\documentstyle[11pt]{article}
\input{sect}
\begin{document}
\title{Gauge independent critical exponents for QED coupled to a four fermi
interaction with and without a Chern Simons term.}
\author{J.A. Gracey, \\ Department of Applied Mathematics and Theoretical
Physics, \\ University of Liverpool, \\ P.O. Box 147, \\ Liverpool, \\
L69 3BX, \\ United Kingdom.}
\date{}
\maketitle
\vspace{5cm}
\noindent
{\bf Abstract.} We apply the self consistency method for determining critical
exponents to a model with a four fermi interaction coupled to QED and compute
various gauge independent exponents in arbitrary dimensions in the large $N$
expansion at $O(1/N)$. The formalism is developed to include a Chern Simons
term in three dimensions and the effect such a term has on the exponents is
deduced.

\vspace{-16cm}
\hspace{10cm}
{\bf {LTH-300}}
\newpage
\sect{Introduction.}
The self consistency technique to determine critical exponents which was
developed in \cite{1,2} has proved to be extremely successful in obtaining
information about various quantum field theories in the large $N$ expansion to
orders well beyond those obtained by conventional renormalization. Briefly the
method involves exploiting the symmetry properties of the field theory at its
$d$-dimensional critical point defined as the non-trivial zero of the
$\beta$-function. There the theory will be finite since it is a fixed point of
the renormalization group and further it will possess a conformal symmetry. In
other words the fields are massless and, for example, a coordinate space
bosonic propagator would take the form $1/(x^2)^\alpha$ where $\alpha$ is the
critical exponent associated with the field, \cite{1,2}. The basic tenet of the
procedure is then to use this most general form for the propagators and solve
for the anomalous part of $\alpha$ which in effect contains information on the
quantum properties of the model. For instance, the universality principle of
statistical mechanics implies that $\alpha$ will depend on the dimension of
spacetime, $d$, and any internal parameters of the theory, \cite{3}. For
example, if a field belongs to an $N$-dimensional multiplet then the exponents
will be a function of $N$. By substituting the general form of the propagator
into the Schwinger Dyson equations truncated to the particular order one is
working to in $1/N$, if $N$ is large, which are valid in the critical region,
one obtains a set of self consistent equations whose solution yields a
(closed) analytic expression as a function of $d$ for the anomalous piece.
Using a renormalization group analysis \cite{3} the expression for the
anomalous part of an exponent can be related to the renormalization group
functions, such as the $\beta$-function, at criticality. In previous work we
were primarily interested in using these analytic expressions to gain new
information on the perturbative series of the renormalization group functions
at all orders in the perturbative coupling constant, \cite{4,5}. In particular
the critical exponents corresponding to the $\beta$-function, $\lambda$, have
been computed to $O(1/N^2)$ for the bosonic $O(N)$ $\sigma$ model in \cite{2}
and its supersymmetric extension in \cite{5}. More recently, the methods have
been extended to include fermions. This was carried out first for the $O(N)$
Gross Neveu model in \cite{4} in order to understand how matter fields could
be included in the formalism. These results were then extended to QED with
$N_{\! f}$ flavours of electrons, \cite{6}, and the electron anomalous
dimension, $\eta$, was computed at $O(1/N^2)$ in the Landau gauge in
arbitrary dimensions in \cite{7}, which therefore included new results on
the three dimensional model. Also, $\lambda$ has been computed within the
self consistency approach at $O(1/N)$, \cite{8,9}.

In this paper, we will consider a model which is an amalgam of QED and the
$O(N)$ Gross Neveu model which is equivalent to a four fermi interaction. In
particular we will compute the $\lambda$ critical exponents which in a single
coupling model would relate to the critical slope of the $\beta$-function and
in a model with a gauge field these are gauge independent. The motivation for
considering such a model is primarily its relation to various physical
phenomena. For instance, several authors have recently considered the three
dimensional model, for which we will derive new results as a corollary of the
arbitrary dimensional result, for its relation to high $T_c$ superconductivity,
\cite{10}. Clearly, one requires theoretical predictions from such models in
order to compare with experimental data and in particular there is a need to
derive gauge independent exponents since these can be measured physically.
Although the $\lambda$ exponent has been determined in pure QED, \cite{9},
within the framework of \cite{2}, that calculation was in the Landau gauge. So
a second aim of our presentation is to demonstrate the gauge independence of
the results not only for QED but also for the amalgam model. Other related work
includes the determination of the value of $N$ for which chiral symmetry is
restored in the three dimensional model, which was discussed in \cite{11} and
requires the exact evaluation of several exponents. Our work will lay the
foundation for improving the accuracy to higher orders.

A further motivation is to understand the self consistency formalism itself in
the direct computation of $\lambda$ exponents in models with more than one
coupling constant and equally to try and ascertain what, if any, are its
limitations. This is far from being an academic exercise since it is important
to extend the formalism to more realistic models which have more than one
coupling constant. It will turn out that the model we discuss provides an
excellent testing ground to fully understand these points.

Finally, as our model will involve a $U(1)$ gauge field we can restrict our
analysis to three dimensions and introduce a Chern Simons term \cite{A,B} to
determine the consequences such a topological term has on the gauge
independent exponents. For instance, knowledge of the $\theta$ dependence at
leading order, where $\theta$ is the coupling of the Chern Simons term, means
that one could deduce from experiments whether $\theta$ was non-zero or not in
reality. Indeed it is believed that there is an intimate connection between a
Chern Simons term and anyonic superconductivity, \cite{12}, and therefore it is
important to have a comprehensive analysis of the realistic model considered
here. Recently, various authors studied the effect such a term had on the
exponents of the three dimensional bosonic $\sigma$ model on $CP(N)$,
\cite{13,14}. Briefly, non-trivial $\theta$ dependent results were obtained.
Since that model and the one we will consider here are related, by, for
example, being the respective sectors of the supersymmetric $\sigma$ model on
$CP(N)$, we would expect a similar non-trivial $\theta$ dependence for the
gauge independent exponents here. Indeed the self consistency method employed
here has one distinct advantage over conventional perturbative techniques in
models with an anti-symmetric tensor which therefore includes Chern Simons
theories. Since the dimension of spacetime remains fixed in the programme of
\cite{1,2}, one does not have to extend the definition of these quantities to
arbitrary dimensions when using dimensional regularization and hence the
subtle ambiguities which then might occur will be avoided.

The paper is organized as follows. We introduce the amalgam model in section 2
and review briefly the leading order method to determine the fermion anomalous
dimension. In section 3 the corrections to this formalism are included and the
consistency equation to determine the $\lambda$ exponents are given. The rather
technical discussion of how to evaluate the two loop massless Feynman diagrams
which arise is given in section 4, whilst we derive arbitrary dimensional
values for the exponents in section 5. The inclusion of the Chern Simons term
is presented in section 6 and we conclude with a discussion in section 7.
Various appendices contain lists of integrals which are needed to evaluate the
graphs.

\sect{Review of basic formalism.}
We begin by recalling the basic features of the model we will be concerned with
as well as reviewing briefly the previous application of the techniques of
\cite{1,2} to the model, \cite{15}, which has the following lagrangian
\begin{equation}
L ~=~ i \bar{\psi}^i \partialslash \psi^i - \frac{(F_{\mu\nu})^2}{4e^2}
+ A_\mu \bar{\psi}^i \gamma^\mu \psi^i + \rho \bar{\psi}^i \psi^i
- \frac{\rho^2}{2g^2} - \frac{(\partial_\mu A^\mu)^2}{2be^2}
\end{equation}
where we have rescaled the coupling constant, $e$, of the $U(1)$ gauge field
$A_\mu$ into the terms quadratic in $A_\mu$ in preparation for applying the
method of \cite{1,2}. We have included a covariant gauge fixing term with
parameter $b$. The four fermi interaction can be obtained explicitly by
eliminating the auxiliary field $\rho$ but we use the form (2.1) as the method
we use to evaluate massless Feynman diagrams is more easily applied to models
with $3$-vertices. Finally, $\psi^i$ is the fermion field where $1$ $\leq$ $i$
$\leq$ $N$ and we will use $1/N$ as our expansion parameter and assume $N$ is
large.

We now illustrate the basic ideas discussed in the introduction and as a
consequence determine the leading order anomalous dimension of $\psi^i$ for
an arbitrary covariant gauge which has not been given before. The first step is
to write down the forms of the respective propagators of (2.1) in the critical
region where they obey asymptotic scaling, consistent with Lorentz symmetry.
Thus, as $x$ $\rightarrow$ $0$, \cite{1,4,7,16},
\begin{eqnarray}
\psi^i(x) & \sim & \frac{A\xslash}{(x^2)^\alpha} ~~~~,~~~~
\rho(x) ~ \sim ~ \frac{B}{(x^2)^\beta} \nonumber \\
A_{\mu\nu}(x) & \sim & \frac{C}{(x^2)^\gamma} \left[ \eta_{\mu\nu}
+ \frac{2\gamma(1-b)}{(2\mu-2\gamma-1+b)} \frac{x_\mu x_\nu}{x^2} \right]
\end{eqnarray}
where $A$, $B$ and $C$ are amplitudes which are independent of $x$, $\mu$ $=$
$d/2$ and $\alpha$, $\beta$ and $\gamma$ are the respective critical exponents
of the fields. We define their anomalous pieces by
\begin{equation}
\alpha ~=~ \mu + \half \eta ~~,~~ \beta ~=~ 1 - \eta - \chi_\rho ~~,~~
\gamma ~=~ 1 - \eta - \chi_A
\end{equation}
where $\eta$ is the anomalous dimension of the fermion and $\chi_\rho$ and
$\chi_A$ are the respective anomalous dimensions of the $3$-vertices of (2.1).
In (2.2) we have included the gauge parameter $b$. Previously we were
interested in exponents solely in the Landau gauge and as part of our aim is to
demonstrate the gauge independence of certain exponents we need to compute not
only $\eta$, $\chi_\rho$ and $\chi_A$ as functions of $b$ but also the various
combinations of the amplitudes. To ease later computations of the $2$-loop
graphs we have chosen to work in coordinate space. Of course when dealing with
massless fields there is a close correspondence between coordinate and momentum
spaces which is effected through the Fourier transform, \cite{1},
\begin{equation}
\frac{1}{(x^2)^\alpha} ~=~ \frac{a(\alpha)}{2^{2\alpha}\pi^\mu} \int_k
\frac{e^{ikx}}{(k^2)^{\mu-\alpha}}
\end{equation}
valid for all $\alpha$ where $a(\alpha)$ $=$ $\Gamma(\mu-\alpha)/
\Gamma(\alpha)$. With this mapping it is straightforward to transform to
$k$-space to see that it has the more recognisable form for a gauge field in
an arbitrary covariant gauge.

As we will be examining the Dyson equations, we will also require the
asymptotic scaling forms of the $2$-point functions for each field. These
are deduced from the propagator functions by first mapping to momentum space,
carrying out the inversion there before mapping the inverted function back to
coordinate space. For the gauge field, this procedure needs to be modified
slightly in that only the transverse part is physically relevant,
\cite{16,17,18}.  Thus the momentum space inversion is restricted to the
transverse subspace ie
\begin{equation}
\eta_\mu^{~\nu} - \frac{p_\mu p^\nu}{p^2} ~=~ \tilde{A}^{-1}_{\mu\sigma}(p)
\tilde{A}^{\sigma\nu} (p)
\end{equation}
Thus the respective asymptotic forms of the $2$-point functions are,
\cite{1,4,7}
\begin{eqnarray}
\psi^{-1}(x) & \sim & \frac{r(\alpha-1)\xslash}{A(x^2)^{2\mu-\alpha+1}}
{}~~,~~~ \rho^{-1}(x) ~ \sim ~ \frac{p(\beta)}{B(x^2)^{2\mu-\beta}} \nonumber
\\
A^{-1}_{\mu\nu} & \sim & \frac{m(\gamma,b)}{C(x^2)^{2\mu-\gamma}} \left[
\eta_{\mu\nu} + \frac{2(2\mu-\gamma)}{(2\gamma-2\mu-1)} \frac{x_\mu x_\nu}{x^2}
\right]
\end{eqnarray}
where
\begin{eqnarray}
r(\alpha) & = & \frac{\alpha a(\alpha-\mu)}{\pi^{2\mu} (\mu-\alpha)
a(\alpha)} ~~~~,~~~~ p(\beta) ~ = ~ \frac{a(\beta-\mu)}{\pi^{2\mu}
a(\beta)} \nonumber \\
m(\gamma,b) & = & \frac{(2\mu-2\gamma-1+b)(2\mu-2\gamma+1)a(\gamma-\mu)}
{4\pi^{2\mu} (\mu-\gamma)^2 a(\gamma)}
\end{eqnarray}

With (2.2) and (2.7), we now solve for $\eta$ by substituting these for the
lines of the truncated skeleton Dyson equations with dressed propagators given
in fig. 1. The use of dressed propagators is to avoid double counting since
the inclusion of the anomalous dimension in the propagator equates to the
contributions from non-dressed propagators, \cite{1,2}. Thus we represent the
Dyson equations in the critical region respectively by
\begin{eqnarray}
0 &=& r(\alpha-1) ~+~ z ~+~ 2y \left[ \frac{\gamma(1-b)}{(2\mu-2\gamma-1+b)}
- \mu + 1 \right] \nonumber \\
0 &=& p(\beta) ~+~ 4Nz \nonumber \\
0 &=& \frac{(\gamma-\mu)m(\gamma,b)}{(2\gamma-2\mu-1)} ~-~ \frac{4(\alpha-1)Ny}
{(2\alpha-1)}
\end{eqnarray}
where we have defined $z$ $=$ $A^2B$ and $y$ $=$ $A^2C$. The equation
corresponding to $A_\mu$ is obtained by mapping the representation of the Dyson
equation to momentum space and projecting out the physical contribution which
is equivalent to the transverse component, before mapping back to coordinate
space, [16-18]. Since the only unknowns at leading order in (2.8) are $z$, $y$
and $\eta$, then eliminating the former two we have
\begin{equation}
\eta_1 ~=~ \frac{[4\mu^2-10\mu+5 +(2\mu-1)b]\Gamma(2\mu-1)}
{4\Gamma(2-\mu)\Gamma^3(\mu)}
\end{equation}
where we have expanded $\eta$ $=$ $\sum_{i=1}^\infty \eta_i/N^i$. Setting $b$
$=$ $0$ it agrees with the Landau gauge result of \cite{15} which provides a
check on the correctness of (2.9). We also note that in three dimensions
\begin{equation}
\eta_1 ~=~ \frac{2(2b-1)}{\pi^2}
\end{equation}
For later purposes we also record
\begin{equation}
z_1 ~=~ - \, \frac{\Gamma(2\mu-1)}{4\pi^{2\mu}\Gamma(\mu-1) \Gamma(1-\mu)}
{}~,~~ y_1 ~=~ - \, \frac{(2\mu-3+b)\Gamma(2\mu)}{16\pi^{2\mu}\Gamma(2-\mu)
\Gamma(\mu)}
\end{equation}

Whilst this completes the leading order analysis of the formalism we will
use it is worthwhile recording that $\eta$ has been computed to $O(1/N^2)$ in
this model in the Landau gauge. For completeness we note, \cite{19},
\begin{eqnarray}
\eta_2 &=& \frac{(2\mu-1)S^2_d}{4\mu(\mu-1)^2} \left[ 4(\mu-1)^2\hat{\Psi}(\mu)
+ 16\mu^2(\mu-4) + 11(9\mu-5) + \frac{4}{(\mu-2)} \right. \nonumber \\
&+& \left. 3\mu(\mu-1)(4\mu^2-10\mu+5) \left( \hat{\Theta}(\mu)
- \frac{1}{(\mu-1)^2} - \frac{(2\mu-1)}{3\mu(\mu-1)} \right) + \frac{8}{\mu}
\right] \nonumber \\
\end{eqnarray}
where $S_d$ $=$ $a(2\mu-2)a^2(1)/\Gamma(\mu)$, $\hat{\Psi}(\mu)$ $=$
$\psi(2\mu-2)$ $-$ $\psi(1)$ $+$ $\psi(3-\mu)$ $-$ $\psi(\mu)$ and
$\hat{\Theta}(\mu)$ $=$ $\psi^\prime(\mu)$ $-$ $\psi^\prime(1)$ with
$\psi(\mu)$ the logarithmic derivative of the $\Gamma$-function. Also,
$\chi_\rho$ and $\chi_A$ have been given at $O(1/N)$ in \cite{15} by
renormalizing the $3$-vertices in an extension of the methods of \cite{16,20}.
It is straightforward to extend \cite{15} to include $b$ $\neq$ $0$ to find
\begin{equation}
\chi_{\rho \, 1} ~=~ - \, \frac{[4\mu^2-6\mu+3+(2\mu-1)b] \Gamma(2\mu-1)}
{4\Gamma^3(\mu) \Gamma(2-\mu)}
\end{equation}
and $\chi_{A \, 1}$ $=$ $-$ $\eta_1$ at leading order. The latter relation is
consistent with the Ward identity of the model which implies that $\gamma$ $=$
$1$ at all orders in large $N$, \cite{7,15}.

\sect{Corrections to scaling functions.}
In this section we introduce the formalism for obtaining the gauge independent
critical exponents we are interested in. It involves considering corrections to
the asymptotic scaling forms of the previous section, where the (unknown)
exponents are related to the slope of the critical $\beta$-function. The form
of these functions have been applied in other models such as the bosonic
$O(N)$ $\sigma$ model \cite{2}, the $O(N)$ Gross Neveu model \cite{4} and QED
\cite{9}. However, each of these involved a single coupling constant unlike
(2.1). Therefore we need to incorporate corrections for both couplings which
then take the form
\begin{eqnarray}
\psi(x) & \sim & \frac{A\xslash}{(x^2)^\alpha} [ 1 + A^\prime
(x^2)^{\lambda_g} + A^{\prime\prime} (x^2)^{\lambda_e} ] \nonumber \\
\rho(x) & \sim & \frac{B}{(x^2)^\beta} [ 1 + B^\prime (x^2)^{\lambda_g}
+ B^{\prime\prime}(x^2)^{\lambda_e} ] \nonumber \\
A_{\mu\nu}(x) & \sim & \frac{C}{(x^2)^\gamma} \left[ \eta_{\mu\nu}
+ \frac{2\gamma(1-b)}{(2\mu-2\gamma-1+b)} \frac{x_\mu x_\nu}{x^2} \right.
\nonumber \\
&+& \left. C^\prime (x^2)^{\lambda_g} \left( \eta_{\mu\nu}
+ \frac{2(\gamma-\lambda_g)(1-b)}{(2\mu-2\gamma+2\lambda_g-1+b)}
\frac{x_\mu x_\nu}{x^2} \right) \right. \nonumber \\
&+& \left. C^{\prime \prime} (x^2)^{\lambda_e} \left( \eta_{\mu\nu}
+ \frac{2(\gamma-\lambda_e)(1-b)}{(2\mu-2\gamma+2\lambda_e-1+b)}
\frac{x_\mu x_\nu}{x^2} \right) \right]
\end{eqnarray}
where $\lambda_g$ $=$ $\mu$ $-$ $1$ $+$ $\sum_{i=1}^\infty
\lambda_{g \, i}/N^i$, $\lambda_e$ $=$ $\mu$ $-$ $2$ $+$ $\sum_{i=1}^\infty
\lambda_{e \, i}/N^i$ are the respective exponents corresponding to the
$\beta$-function slope at criticality and their leading order terms are
related to the canonical dimensions of the couplings. The new amplitudes
$A^\prime$, $A^{\prime\prime}$ etc are the associated $x$-independent
amplitudes. The asymptotic scaling forms of the $2$-point function are given
by the same procedure as the previous section and we find
\begin{eqnarray}
\psi^{-1}(x) & \sim & \frac{r(\alpha-1)\xslash}{A(x^2)^{2\mu-\alpha+1}}
[ 1 - A^\prime s(\alpha-1, \lambda_g)(x^2)^{\lambda_g}
- A^{\prime\prime} s(\alpha-1, \lambda_e) (x^2)^{\lambda_e} ] \nonumber \\
\rho^{-1}(x) & \sim & \frac{p(\beta)}{B(x^2)^{2\mu-\beta}} [ 1
- B^\prime q(\beta,\lambda_g)(x^2)^{\lambda_g} - B^{\prime\prime}
q(\beta,\lambda_e) (x^2)^{\lambda_e} ] \nonumber \\
A^{-1}_{\mu\nu}(x) & \sim & \frac{m(\gamma,b)}{C(x^2)^{2\mu-\gamma}}
\left[ \eta_{\mu\nu} + \frac{2(2\mu-\gamma)}{(2\gamma-2\mu-1)}
\frac{x_\mu x_\nu}{x^2} \right. \nonumber \\
&-& \left. C^\prime n(\gamma,\lambda_g,b) (x^2)^{\lambda_g} \left(
\eta_{\mu\nu} + \frac{2(2\mu-\gamma-\lambda_g)}{(2\gamma+2\lambda_g-2\mu-1)}
\frac{x_\mu x_\nu}{x^2} \right) \right. \nonumber \\
&-& \left. C^{\prime\prime}n(\gamma,\lambda_e,b)(x^2)^{\lambda_e} \left(
\eta_{\mu\nu} + \frac{2(2\mu-\gamma-\lambda_e)}{(2\gamma+2\lambda_e-2\mu
-1)} \frac{x_\mu x_\nu}{x^2} \right) \right]
\end{eqnarray}
where the various new functions are defined by
\begin{eqnarray}
s(\alpha,\lambda) &=& \frac{\alpha(\alpha-\mu) a(\alpha-\mu+\lambda)
a(\alpha-\lambda)}{(\alpha-\lambda)(\alpha-\mu+\lambda)a(\alpha-\mu)
a(\alpha)} \nonumber \\
q(\beta,\lambda) &=& \frac{a(\beta-\mu+\lambda)a(\beta-\lambda)}
{a(\beta-\mu)a(\beta)} \\
n(\gamma,\lambda,b) &=& \frac{(\mu-\gamma+\lambda)(2\mu-2\gamma-1+b)
(2\mu-2\gamma-2\lambda+1)}{(\mu-\gamma-\lambda)(2\mu-2\gamma+1)
(2\mu-2\gamma+2\lambda-1+b)} \nonumber \\
&\times& \frac{a(\gamma+\lambda-\mu) a(\gamma-\lambda)}{a(\gamma)a(\gamma-\mu)}
\nonumber
\end{eqnarray}
It is worth noting that these corrections do not alter the usual momentum
space structure for covariant gauges.

The next step is to derive a consistency equation analogous to the set (2.8)
whose solution will yield $\lambda_{g \, 1}$ and $\lambda_{e \, 1}$.
Ordinarily these are obtained by substituting the forms (3.1) and (3.2)
into the graphs of fig. 1 as we did in the previous section. However, this
straightforward substitution only works when one has a bosonic model, \cite{2}.
When the fundamental field is purely fermionic then there is what we term a
reordering of graphs which contribute to the $\lambda$ exponents. This was
discussed in \cite{9} but for completeness we recall the salient features
here. When one substitutes for the lines of the graphs of fig. 1 one obtains
a set of three equations which involve the two correction terms. As these
involve different powers of $x^2$ one can decouple the equations into three
separate sets which include (2.8) and one each which determine $\lambda_g$
and $\lambda_e$. For the latter set they can be encoded in a $3\times3$
matrix with the independent amplitudes acting as the basis vectors. To ensure
this forms a consistent set of equations the determinant of this matrix must
vanish and since this involves the one unknown $\lambda$ its solution yields
an arbitrary dimensional expression for $\lambda$. However, unlike the
purely bosonic case of \cite{2} each element of the matrix is not of the same
order in $1/N$. For instance, at $\alpha$ $=$ $\mu$ and $\lambda_g$ $=$ $\mu$
$-$ $1$,
\begin{equation}
s(\alpha-1,\lambda_g) ~=~ \frac{4(\mu-2)}{\mu\eta}
\end{equation}
which is $O(N)$. Whilst it multiplies terms of only $O(1/N)$ in the evaluation
of the determinant which come from the one loop graphs of fig. 1, it in fact
yields contributions to $\lambda$ which are of the correct order in $1/N$,
\cite{9}. If one naively carried out the evaluation the wrong result would be
obtained. This is due to the omission of the graphs of fig. 2 where the
internal $\rho$ or $A_\mu$ field is corrected, which within the matrix will
yield contributions of the same order in $1/N$ as the other elements. Only when
the analogous two loop graph was included in say, QED, did the correct critical
exponent emerge, \cite{8,9}. Thus including these extra graphs of fig. 2, we
obtain the decoupled equations for $\lambda_g$ and $\lambda_e$ as
\begin{eqnarray}
0 &=& [1+s(\alpha-1,\lambda)]\left[ z - 2y \left( \mu-1 - \frac{\gamma(1-b)}
{(2\mu-2\gamma-1+b)}\right)\right] A^\prime \nonumber \\
&+& zB^\prime - 2y C^\prime \left[ \mu-1 - \frac{(\gamma-\lambda)(1-b)}
{(2\mu -2\gamma+2\lambda-1+b)}\right] \nonumber \\
0 &=& 8A^\prime + [ 4q(\beta,\lambda)-z\Gamma_1]B^\prime - y\Gamma_2
C^\prime \\
0 &=& \frac{8(2\alpha-\lambda-2)}{(2\alpha-\lambda-1)} A^\prime
+ z \left( \Pi_2 + \frac{\Xi_2}{2(4\alpha+\beta-2\mu-\lambda-2)}\right)
B^\prime \nonumber \\
&+& \left[ \frac{8(\alpha-1)(2\gamma-2\mu-1)(\gamma+\lambda-\mu)
n(\gamma,\lambda,b)}{(2\alpha-1)(\gamma-\mu)(2\gamma+2\lambda-2\mu-1)}
\right. \nonumber \\
&&+ \left. y \left( \Pi_1 + \frac{\Xi_1}{2(4\alpha+\gamma-2\mu-\lambda-2)}
\right) \right] C^\prime \nonumber
\end{eqnarray}
where $\lambda$ is either $\lambda_g$ or $\lambda_e$ and the four two loop
contributions $\Gamma_i$ and $\Pi_{i\,\mu\nu}$ are evaluated at either of
these values as well. The components of $\Pi_{i \, \mu \nu}$ are defined by
\begin{equation}
\Pi_{i \, \mu\nu} ~=~ \Pi_i \eta_{\mu\nu} ~+~ \Xi_i \frac{x_\mu x_\nu}{x^2}
\end{equation}
Again we have only considered the transverse portion of the $A_\mu$ equation
relative to momentum space as it is the only physically relevant piece,
[9,16-18]. It is worth stressing that the two loop corrections to the
fermion self energy do not in fact need to be considered since their
contributions are suppressed by a factor of $1/N$ relative to the $\Gamma_i$
and $\Pi_{i \, \mu \nu}$ terms.

\sect{Computation of graphs.}
To complete our analysis we require the values of the two loop integrals of
fig. 2 where we recall that the exponent of the completely internal propagator
is shifted from $\beta$ or $\gamma$ by $\lambda_g$ or $\lambda_e$. In the
computation of the exponent $\eta$ at $O(1/N^2)$, \cite{19}, similar two loop
graphs with unshifted exponents were considered. Those graphs are, however,
infinite due to infinities which arise from vertex subgraphs when $\chi_\rho$
and $\chi_A$ are zero. In \cite{2,15} a regularization was introduced by
shifting the $\rho$ and $A_\mu$ exponents by an infinitesimal amount $\Delta$
ie $\beta$ $\rightarrow$ $\beta$ $-$ $\Delta$, $\gamma$ $\rightarrow$ $\gamma$
$-$ $\Delta$ and then the divergent terms are absorbed by a vertex
renormalization to leave a set of finite consistency equations to determine
$\eta_2$, (2.12), \cite{19}. For the graphs we consider here the effect of the
shift by $\lambda$ is to render each two loop graph $\Delta$-finite so that we
do not need to introduce formalism to obtain a $\Delta$-finite set of $\lambda$
consistency equations. Paradoxically, however, these completely finite graphs
are much harder to compute than the divergent ones as we require an exact
expression for each. In the original work of \cite{1,2} the main technique used
to evaluate the massless graphs was based on the method of uniqueness
introduced in \cite{21} and to a lesser extent recursion relations which were
determined by an integration by parts rule. The main uniqueness rule for a
bosonic vertex is illustrated in fig. 3 where integrating over the internal
point of the vertex on the left side yields the product of the three
propagators on the right side when the arbitrary exponents, $\alpha_i$, are
constrained to satisfy the uniqueness condition, $\sum_{i=1}^3 \alpha_i$ $=$
$2\mu$, and we set $\nu(\alpha_1,\alpha_2,\alpha_3)$ $=$ $\pi^\mu \prod_{i=1}^3
a(\alpha_i)$. This rule was used extensively in \cite{2} to obtain $\lambda$
for the bosonic $O(N)$ $\sigma$ model.

The situation involving fermions is somewhat more involved, \cite{4,9}. In
computing $\Pi_{1 \, \mu\nu}$ for QED \cite{4} or $\Gamma_1$ for the Gross
Neveu model \cite{9} we had to adapt this strategy somewhat. For a graph with
a completely internal gauge field we first had to integrate by parts to
simplify the contribution from the $x_\mu x_\nu$ term of (2.2). Then the trace
over the $\gamma$-matrices of the numerator of the integral representing the
graph was performed. Consequently the original graph is obtained as a sum of
purely bosonic two loop graphs where the exponents of each are related by
addition or subtraction of integers to the corresponding exponents of the
original graph \cite{4}. Whilst each of these constituent graphs is
$\Delta$-finite they are not finite when one substitutes the leading order
values of $\alpha$ $=$ $\mu$ and $\beta$ $=$ $\gamma$ $=$ $1$. The reason for
this is simple. When the exponent of a bosonic line is $\mu$, which is its
anti-uniqueness value \cite{1,2}, it corresponds to a divergence of the form
$a(\mu)$, see (2.4), which is a different infinity to those needing a
regularization by $\Delta$. These other infinities are naturally controlled
by the anomalous part of the fermion exponent, $\eta/2$. In order to extract
a finite value for each graph it is best to write the constituent $2$-loop
graphs in such a way that the infinities which enter as $1/\eta$ cancel
naturally among themselves. One way to achieve this is to write each
constituent graph as far as is possible in a form which has the factors
$(\mu-\alpha)^{-1}$ explicit and which multiply a graph which is completely
finite. The form of such a graph will be one where the exponent of each line
corresponding to a fermion is $\alpha$ $-$ $1$, or $\alpha$ $-$ $2$, which are
not anti-uniqueness values. To achieve this we apply various recursion
relations for massless $2$-loop Feynman graphs which were discussed in
\cite{9} and used also in \cite{19}. In the QED work of \cite{9} several
finite linear combinations were obtained but it turns out that more are
required for (2.1).

To illustrate these points we will discuss the computation of $\Gamma_1$
where for the moment the exponent of the $\rho$ propagator will be left as a
free parameter, $\xi$, but will be set to either $2$ $-$ $\mu$ or $3$ $-$ $\mu$
at the end of the calculation. With the trace convention $\mbox{tr}1$ $=$ $4$,
we rewrite $\Gamma_1$ as
\begin{equation}
4\la \alpha-1, \alpha, \alpha-1, \alpha, \xi \ra
\end{equation}
where the general two loop self energy massless Feynman graph is illustrated in
fig. 4 and defined as $\la \alpha_1, \alpha_2, \alpha_3, \alpha_4, \alpha_5
\ra$ where $\alpha_i$ are arbitrary exponents. In fig. 4, the top and bottom
vertices are vertices of integration. In \cite{9} we defined the linear
combination (4.1) by $2 A_\xi$ and it will turn out that it is finite. To see
this we apply the recursion relation (A.1) to the second graph of $A_\xi$ in
(4.1) to give
\begin{eqnarray}
&& \frac{2(\mu-\alpha)}{(\alpha-1)} \la \alpha, \alpha-1, \alpha, \alpha-1,
\xi \ra \nonumber \\
\la \alpha, \alpha-1, \alpha-1, \alpha, \xi \ra
\end{eqnarray}
Next (A.2) is applied to the first term of (4.2) and after some rearrangement
we have
\begin{eqnarray}
&& \frac{[2(\mu-\alpha)(\mu-\alpha-\xi)+\xi(\xi-\mu+1)]}
{(\alpha-1)^2} \la \alpha-1, \alpha, \alpha, \alpha-1, \xi \ra \nonumber \\
\alpha-1, \xi+1 \ra
\end{eqnarray}
Finally, (A.3) is applied to the first graph of (4.3), and after some algebra
yields
\begin{eqnarray}
&& \frac{2(\mu-\alpha)(2\mu-2\alpha-\xi)(2\alpha+\xi-\mu-1)}{(\alpha-1)^3} \la
\alpha-1, \alpha-1, \alpha, \alpha-1, \xi+1 \ra \nonumber \\
&&+~~ M(\alpha,\xi) \frac{(4\alpha+\xi-2\mu-3)}{(\alpha-1)^4}
\la \alpha-1, \alpha-1, \alpha-1, \alpha-1, \xi+1 \ra
\end{eqnarray}
where $M(\alpha,\xi)$ $=$ $(3\mu-4\alpha-\xi+2)$$[2(\alpha-1)(\mu-\alpha-\xi)$
$-$ $(2\mu-2\alpha-\xi)(2\alpha+\xi-\mu-1)]$. Thus we have written $A_\xi$ in
terms of two graphs. The first is finite when $\alpha$ $=$ $\mu$ and $\xi$ is
set to either $2$ $-$ $\mu$ or $3$ $-$ $\mu$, whilst the second term has all
but one exponent at $\alpha$ $-$ $1$. A factor $(\mu-\alpha)$ premultiplies
this term, however, and it cancels the implicit $(\mu-\alpha)^{-1}$ singularity
of the graph. In \cite{9} an explicit form for this graph was given and rather
than rederive this result here, we merely state it, ie
\begin{eqnarray}
&& b(\alpha,\xi) \la \alpha-1, \alpha-1, \alpha-1, \alpha, \xi \ra \nonumber \\
&&~=~ c(\alpha,\xi) \la \alpha-1, \alpha-1, \alpha-1, \alpha-1, \xi+1 \ra
\nonumber \\
&&~+~ \frac{(\xi+1)d(\alpha,\xi)}{(\mu-\alpha)} \la \alpha-1, \alpha-3,
\alpha-1, \alpha-1, \xi+2 \ra \nonumber \\
&&~+~ \frac{(2\alpha-\mu-3)d(\alpha,\xi)}{(2\alpha+\xi-\mu-1)(2\mu-2\alpha
-\xi)} \la \alpha-1, \alpha-1, \alpha-2, \alpha-1, \xi+1 \ra \nonumber \\
&&~+~ \frac{(\xi+1)(2\alpha-\mu-3)}{(\mu-\alpha)(\alpha-1)}
\la \alpha-1, \alpha-2, \alpha-1, \alpha-1, \xi+2 \ra
\end{eqnarray}
where
\begin{eqnarray}
b(\alpha,\xi) &=& 1 - \frac{(2\alpha-\mu-3)(\alpha-2)}{(2\alpha+\xi-\mu-2)
(2\alpha+\xi-\mu-1)(2\mu-2\alpha-\xi)} \nonumber \\
c(\alpha,\xi) &=& \frac{(\mu-\alpha-\xi)}{(\alpha-1)}
+ \frac{(2\alpha-\mu-3)(\alpha-2)(2\mu-2\alpha-\xi+1)}{(\alpha-1)
(2\alpha+\xi-\mu-1)(2\mu-2\alpha-\xi)} \nonumber \\
d(\alpha,\xi) &=& \frac{(3\mu-4\alpha-\xi+3)(4\alpha-2\mu+\xi-4)}
{(\alpha-1)(2\alpha+\xi-\mu-2)} \nonumber
\end{eqnarray}
and we  have displayed the $1/(\mu-\alpha)$ pieces explicitly. So we have
obtained an expression, (4.4), albeit involved, which is completely finite at
$\alpha$ $=$ $\mu$, $\beta$ $=$ $\gamma$ $=$ $1$ and all that remains is to
evaluate it. In appendix B we have noted the values of several of the graphs
which are needed for (4.4) and other combinations. They have been derived using
the uniqueness rule of fig. 3 and recursion relations similar to (A.1)-(A.3).
We find
\begin{equation}
A_{2-\mu} ~=~ \frac{2\pi^{2\mu}}{(\mu-1)^2 \Gamma^2(\mu)} ~~~,~~~
A_{3-\mu} ~=~ - \, \frac{2\pi^{2\mu} (\mu^2-4\mu+2)}{(\mu-1)^2(\mu-2)
\Gamma^2(\mu)}
\end{equation}
where we have used
\begin{eqnarray}
[(\mu-\alpha) \la \alpha, \alpha-1, \alpha-1, \alpha-1, 3-\mu \ra]
|_{\alpha=\mu} &=& \frac{\pi^{2\mu}(\mu-1)}{(\mu-2) \Gamma^2(\mu)} \nonumber \\
{[} (\mu-\alpha)\la \alpha, \alpha-1, \alpha-1, \alpha-1, 4-\mu \ra {]}
|_{\alpha=\mu} &=& \frac{\pi^{2\mu}(\mu-1)}{2(\mu-3)\Gamma^2(\mu)}
\end{eqnarray}
This in effect completes the evaluation of $\Gamma_1$ at both leading order
values of $\lambda$ we are considering and we devote the remainder of this
section to the essential points in determining the other graphs.

First, we define the other finite linear combinations of graphs analogous
to $A_\xi$ which arise when the constituent graphs are rewritten in
bosonic components as
\begin{eqnarray}
B_\xi &=& \la \alpha, \alpha, \alpha, \alpha, \xi \ra + 2 \la \alpha,
\alpha-1, \alpha-1, \alpha, \xi \ra \nonumber \\
&-& 4 \la \alpha-1, \alpha, \alpha, \alpha, \xi \ra
+ \la \alpha, \alpha, \alpha, \alpha, \xi-1 \ra \\
C_\xi &=& \la \alpha-1, \alpha, \alpha, \alpha-1, \xi-1 \ra
- \la \alpha, \alpha, \alpha-1, \alpha-1, \xi \ra \\
D_\xi &=& 2 \la \alpha-1, \alpha, \alpha-2, \alpha, \xi \ra
- \la \alpha-1, \alpha, \alpha-1, \alpha, \xi-1 \ra \nonumber \\
&-& \la \alpha-1, \alpha, \alpha-1, \alpha, \xi \ra \\
E_\xi &=& \la \alpha-1, \alpha-1, \alpha, \alpha, \xi \ra
- 2\la \alpha-1, \alpha-1, \alpha-1, \alpha-1, \xi \ra
\end{eqnarray}
These have been evaluated for some values of $\xi$ in terms of finite graphs
in \cite{9} and we note that their explicit forms have been given in
appendix B. For completeness we note that their values at $\alpha$ $=$
$\mu$ are
\begin{eqnarray}
B_{2-\mu} &=& \frac{2(2\mu-3)\pi^{2\mu}}{(\mu-1) \Gamma^2(\mu)} ~~~,~~~
B_{3-\mu} ~=~ \frac{(2\mu^2-7\mu+4)\pi^{2\mu}}{(\mu-1)(\mu-2)\Gamma^2(\mu)}
\nonumber \\
C_{3-\mu} &=& \frac{2(\mu-3)\pi^{2\mu}}{(\mu-2)\Gamma^2(\mu)} ~~,~~
D_{3-\mu} ~=~ E_{3-\mu} ~=~ - \, \frac{2\pi^{2\mu}}{\Gamma^2(\mu)}
\end{eqnarray}
In several graphs, however, it turns out that we have to consider the
additional combinations $B_{\xi-1}$ $-$ $2C_\xi$ and $C_\xi$ $+$ $D_\xi$
$+$ $E_\xi$ which are similarly written as a sum of finite integrals through
recursion relations. The need to consider these combinations is due to the
appearance of the accidental singularity at $2\alpha$ $+$ $\xi$ $-$ $\mu$
$-$ $1$ $=$ $0$, which occurs at $\xi$ $=$ $1$ $-$ $\mu$ and arises in
several graphs involving $\lambda_g$. For example, after extensive use of
recursion relations we have
\begin{eqnarray}
B_{\xi-1} - 2C_\xi &=& P(\xi) \la \alpha-1, \alpha-1, \alpha-1, \alpha-1,
\xi+2 \ra \nonumber \\
&+& Q(\xi) \la \alpha-1, \alpha-1, \alpha, \alpha-1, \xi+2 \ra
\end{eqnarray}
where
\begin{eqnarray}
P(\xi) &=& \frac{(3\mu-4\alpha-\xi+1)(4\alpha+\xi-2\mu-2)}
{(\alpha-1)^3(2\mu-2\alpha-\xi)} \nonumber \\
&\times& \left[ (2\mu-2\alpha-\xi-1)(2\alpha+\xi-\mu) \left(
\frac{(\mu-1)}{(\alpha-1)} + \frac{2(\mu-\alpha)}{(2\mu-2\alpha-\xi)}\right)
\right. \nonumber \\
&-& \left. 2(\mu-1)(\mu-\alpha-\xi-1) \frac{}{} \right] \nonumber \\
Q(\xi) &=& \frac{2(\mu-\alpha)}{(\alpha-1)(2\mu-2\alpha-\xi)} \left[ 2(1-\mu)
\right. \nonumber \\
&+& \left. 2(2\alpha+2\xi-\mu+1)\left( \frac{\mu-1}{\alpha-1}
+ \frac{2(\mu-\alpha)}{2\mu-2\alpha-\xi}\right) \right. \nonumber \\
&-& \left. (\xi+1)(\mu+\xi)\left( \frac{1}{(\alpha-1)} + \frac{2}{(2\mu-2\alpha
-\xi)} \right) \right. \nonumber\\
&-& \left. \frac{(2\mu-2\alpha-\xi-1)(2\alpha+\xi-\mu)}{(\alpha-1)} \left(
\frac{(\mu-1)}{(\alpha-1)} + \frac{2(\mu-\alpha)}{(2\mu-2\alpha-\xi)} \right)
\right. \nonumber \\
&-& \left. \frac{(\mu-\xi-2)(\xi+1)}{(\alpha-1)} + \frac{2(\mu-1)(\mu-\alpha
-\xi-1)}{(\alpha-1)} \right]
\end{eqnarray}
Also,
\begin{eqnarray}
&& C_{\xi+1} + D_{\xi+1} + E_{\xi+1} \nonumber \\
&&~=~ \frac{(\xi+1)(\mu-\xi-2)(2\alpha+\xi-2)}{(\alpha-1)^2(2\mu-2\alpha-\xi)}
\la \alpha-1, \alpha-1, \alpha-1, \alpha-1, \xi+2 \ra \nonumber \\
&&~-~ \frac{4(\mu-\alpha)(2\alpha+\xi-\mu-1)}{(\alpha-1)(2\mu-2\alpha-\xi)}
\la \alpha-1, \alpha-1, \alpha, \alpha-1, \xi+1 \ra
\end{eqnarray}
which requires the use of
\begin{eqnarray}
&& \la \alpha-1, \alpha-1, \alpha-1, \alpha, \xi+1 \ra \nonumber \\
&&~=~ \frac{(2\alpha+\xi-\mu)}{(2\mu-2\alpha-\xi)} \la \alpha-1, \alpha-1,
\alpha, \alpha-1, \xi+2 \ra \nonumber \\
&&~-~ \frac{(4\alpha+2\xi-3\mu)}{(2\mu-2\alpha-\xi)} \la \alpha-1, \alpha-1,
\alpha-1, \alpha-1, \xi+2 \ra \nonumber \\
&&~-~ \frac{(4\alpha+2\xi-3\mu)}{(2\mu-2\alpha-\xi-1)} \la \alpha-1, \alpha-2,
\alpha, \alpha-1, \xi+2 \ra
\end{eqnarray}
and
\begin{eqnarray}
&& R(\xi) \la \alpha-1, \alpha, \alpha-2, \alpha-1, \xi+2 \ra \nonumber \\
&&~=~ \la \alpha-1, \alpha-1, \alpha, \alpha-1, \xi+2 \ra \nonumber \\
&&~-~ \frac{(\xi+2)(\mu-\xi-3)}{(\alpha-1)(\alpha-2)} \la \alpha-2, \alpha-1,
\alpha-1, \alpha-1, \xi+3 \ra \nonumber \\
&&~-~ \frac{(2\alpha-\mu-2)}{(\alpha-1)} \la \alpha-1, \alpha-1, \alpha-1,
\alpha-1, \xi+2 \ra
\end{eqnarray}
to evaluate it at $\xi$ $=$ $1$ $-$ $\mu$, where
\begin{equation}
R(\xi) ~=~ \frac{(2\alpha-\mu-2)(2\alpha+\xi-\mu-1)(2\mu-2\alpha-\xi)}
{(2\alpha+\xi-\mu)(2\mu-2\alpha-\xi-1)(\alpha-2)} \nonumber
\end{equation}
Thus with (4.16) and (4.17), which imply
\begin{equation}
\!\! \left. \frac{}{} [ (\mu-\alpha)(2\alpha+\xi-\mu-1)\la \alpha-1, \alpha-1,
\alpha, \alpha-1, \xi+1 \ra ] \right|_{\mbox{\footnotesize ${\begin{array}{l}
                \alpha = \mu \\
                \xi = 1 - \mu
               \end{array}}$}}
\end{equation}
we have
\begin{eqnarray}
B_{1-\mu} - 2 C_{2-\mu} &=& \frac{2\pi^{2\mu}}{(\mu-1)^2\Gamma^2(\mu)}
\nonumber \\
C_{2-\mu} + D_{2-\mu} + E_{2-\mu} &=& \frac{3\pi^{2\mu}\hat{\Theta}(\mu)}
{\Gamma^2(\mu)} - \frac{4\pi^{2\mu}}{(\mu-1)^2\Gamma^2(\mu)}
\end{eqnarray}

Finally we complete this section by recording the location of these
combinations in the graphs of fig. 2. First, though we note that for graphs
involving an internal gauge field an integration by parts rule is used to
rewrite the $A_\mu \bar{\psi} \gamma^\mu \psi$ vertex within the graph which
has the consequence that the trace over the $\gamma$-matrices is simplified.
In \cite{9} we gave the relevant rule for the Landau gauge but since we are
working in a general covariant gauge we use the results of fig. 5 where the
exponents $\alpha$, $\beta$ and $\gamma$ are arbitrary, the long dashed
line corresponds to the vector $z^\mu$ if $z$ is the location of the internal
vertex of integration and $0$ the origin, and the index $\nu$ in the first
graph of the right side of fig. 5 corresponds to a $\gamma$-matrix,
$\gamma^\nu$. If we denote by $I$ the contribution from the first vertex on
the right side of fig. 5 without the factor and by $II$ the contribution
from the other vertices where both factors are equivalent in the present
situation then we have
\begin{equation}
\Gamma^I_2 ~=~ 4 [ B_{\xi-1} + (\mu-1)A_\xi - 2C_\xi]
\end{equation}
It is easy to see that the factor associated with $\Gamma_2^{II}$ involves
$(\mu-\alpha)^2$, so one does not need to write its bosonic graph in the
same finite combinations, but only isolate the double poles
$(\mu-\alpha)^{-2}$ and evaluate their associated graphs. We note, though,
\begin{eqnarray}
\left. {\frac{8(\mu-\alpha)^2\Gamma^{II}_{2\,\lambda_g}}{(\beta-1-\lambda_g)}}
\right|_{\alpha=\mu} &=& - \, \frac{16\pi^{2\mu}}{(\mu-1)(4\mu-5)\Gamma^2(\mu)}
\nonumber \\
\left. {\frac{8(\mu-\alpha)^2\Gamma^{II}_{2\,\lambda_g}}{(\beta-1-\lambda_e)}}
\right|_{\alpha=\mu} &=& - \, \frac{16\pi^{2\mu}}{(\mu-2)(4\mu-7)\Gamma^2
(\mu)}
\end{eqnarray}
For $\Pi_{i \, \mu\nu}$ considering $\Pi_{i\, \mu}^{~\,~\mu}$ and $\Pi_{i \,
\mu\nu}x^\mu x^\nu$ separately, then
\begin{eqnarray}
\Pi_{2\,\mu}^{~\,~\mu} &=& 4[ B_\xi + (\mu-1)A_\xi] \nonumber \\
\Pi_{2\,\mu\nu}x^\mu x^\nu &=& 2 B_\xi
\end{eqnarray}
Finally,
\begin{eqnarray}
\Pi^{I~\,~\mu}_{1\,\mu} &=& - \, 8(\mu-1)[(\mu-2)A_\xi + B_\xi + B_{\xi-1}
-2C_\xi] \nonumber \\
\Pi_{1 \, \mu\nu}^I x^\mu x^\nu &=& - \, 4[ B_{\xi-1} + 2E_\xi + 2D_\xi
+(\mu-1)B_\xi]
\end{eqnarray}
and $\Pi_{1\,\mu\nu}^{II}$ $=$ $0$ at $\lambda_g$ $=$ $\mu$ $-$ $1$ but
\begin{equation}
\Pi_{1 \, \mu\nu}^{II} ~=~ \frac{16\pi^{2\mu}}{(\mu-2)\Gamma^2(\mu)}
\left[ \eta_{\mu\nu} - \frac{2x_\mu x_\nu}{x^2} \right]
\end{equation}

\sect{Critical exponents.}
In this section we derive expressions for $\lambda_g$ and $\lambda_e$ and
discuss in more detail certain aspects of the formalism. As a preliminary we
note that the values we obtain by assemblying all the components of the
two loop graphs of fig. 2 are first, for $\lambda_g$ $=$ $\mu$ $-$ $1$
\begin{eqnarray}
\Gamma_1 &=& \frac{4\pi^{2\mu}}{(\mu-1)^2\Gamma^2(\mu)} ~~~,~~~
\Gamma_2 ~=~ \frac{16(2\mu-1+b)\pi^{2\mu}}{(4\mu-5+b)(\mu-1)\Gamma^2(\mu)}
\nonumber \\
\Pi_{1 \, \mu\nu} &=& \frac{96\pi^{2\mu}(\mu-1)}{(2\mu-1)(4\mu-5+b)
\Gamma^2(\mu)} \left( \hat{\Theta}(\mu) - \frac{1}{(\mu-1)^2} \right) \left(
\eta_{\mu\nu} - 2\mu \frac{x_\mu x_\nu}{x^2} \right) \nonumber \\
&-& \frac{32(\mu-1)\pi^{2\mu}}{(4\mu-5+b)\Gamma^2(\mu)} \left( \eta_{\mu\nu}
+ 2(\mu-2) \frac{x_\mu x_\nu}{x^2} \right) \nonumber \\
&+& \frac{16(1-b)\pi^{2\mu}}{(4\mu-5+b)(\mu-1) \Gamma^2(\mu)}
\left( \eta_{\mu\nu} - \frac{2x_\mu x_\nu}{x^2} \right) \nonumber \\
\Pi_{2 \, \mu\nu} &=& \frac{4\pi^{2\mu}}{(\mu-1)\Gamma^2(\mu)}
\left( \eta_{\mu\nu} + 2(\mu-2) \frac{x_\mu x_\nu}{x^2} \right)
\end{eqnarray}
and at $\lambda_e$ $=$ $\mu$ $-$ $2$
\begin{eqnarray}
\Gamma_1 &=& - \, \frac{4\pi^{2\mu}(\mu^2-4\mu+2)}{(\mu-1)^2(\mu-2)
\Gamma^2(\mu)} \nonumber \\
\Gamma_2 &=& - \, \frac{16\pi^{2\mu}[(2\mu-3)(\mu^2-5\mu+2)+(\mu-1)(1-b)]}
{(\mu-1)(\mu-2)(4\mu-7+b)\Gamma^2(\mu)} \nonumber \\
\Pi_{1\,\mu\nu} &=& \frac{8\pi^{2\mu}}{(4\mu-7+b)(\mu-1)(\mu-2)(\mu+1)
\Gamma^2(\mu)} \nonumber \\
&\times& \left[ (2\mu-3)(\mu^2-11\mu+16)
\left(\eta_{\mu\nu} - \frac{2x_\mu x_\nu}{x^2} \right) \right. \nonumber \\
&+& \!\! \left. 2(1-b)(\mu-1)\left( \eta_{\mu\nu} - \frac{2x_\mu x_\nu}{x^2}
\right) - 2(2\mu-3)(\mu-2)(\mu-3) \frac{x_\mu x_\nu}{x^2} \right]
\nonumber \\
\Pi_{2 \, \mu \nu} &=& - \, \frac{2\pi^{2\mu}}{(\mu-1)(\mu-2)\Gamma^2(\mu)}
\left[ (\mu-4)\eta_{\mu\nu} - 2\mu(\mu-3)\frac{x_\mu x_\nu}{x^2} \right]
\end{eqnarray}

First, we will derive $\lambda_{g \, 1}$ using the values (5.1). The
determinant which is set to zero to give $\lambda_{g \, 1}$ is given by (3.5).
After factoring out various common pieces, we have $\det M$ $=$ $0$ where $M$
is the matrix
\begin{equation}
\left( {\begin{array}{ccc}
1 & 2 & [(2\mu-3)(2\mu-1)+b] \\
\mu-1 & - (3\mu-2) - \frac{2\tilde{\lambda}_{g \, 1}}{(\mu-1)}
& \frac{(\mu-1)(2\mu-1+b)} {(2\mu-1)} \\
1 & 2 & - \, \frac{8\mu\tilde{\lambda}_{g\,1}}{(2\mu-1)^2}
+ [(2\mu-3)(2\mu-1)+b] \\ \end{array}} \right)
\end{equation}
where $\tilde{\lambda}_{g\,1}$ $=$ $\Gamma(2-\mu)\Gamma^3(\mu)\lambda_{g\,1}/
\Gamma(2\mu-1)$. With several row transformations it is easy to discover that
two solutions for $\lambda_{g \, 1}$ emerge for all $\mu$ ie
\begin{equation}
\lambda_g ~=~ \mu - 1 + O \left( \frac{1}{N^2} \right) ~~~,~~~
\lambda_g ~=~ \mu - 1 - \frac{\Gamma(2\mu+1)}{4\Gamma^3(\mu)\Gamma(2-\mu)N}
\end{equation}
We will disregard the trivial solution on the reasonable grounds that the
inclusion of a QED interaction in a model with a four fermi interaction or
vice versa ought not to cancel the effect of the other to give a trivial
solution in all dimensions. Further, as (5.4) is valid in all dimensions we
can set $d$ $=$ $3$ to find
\begin{equation}
\lambda_g ~=~ \frac{1}{2} ~-~ \frac{12}{\pi^2N}
\end{equation}
It is also important to note that unlike the expressions for $\eta$ the gauge
parameter, $b$, has cancelled in taking the determinant and it is in this
sense that we refer to $\lambda_g$ as being a gauge independent exponent.

Repeating the analysis with (5.2) to determine $\lambda_{e \, 1}$ a different
picture emerges. After several row and column transformations on the
determinant of the analogous $3\times3$ matrix the vanishing of $\det L$ where
$L$ is the following $2\times2$ matrix
\begin{equation}
\left( {\begin{array}{cc}
\frac{2\mu\tilde{\lambda}_{e\,1} + (2\mu-1)(2\mu^2-4\mu+3)}
{(2\mu-3)(\mu-2)} & \frac{(2\mu-1)^2(2\mu-3)(\mu-3)}{(\mu-1)} \\
(\mu^2-1) & \frac{4\mu\tilde{\lambda}_{e\,1}}{(2\mu-1)}
+ \frac{(2\mu-3)(\mu-2)(\mu-3)(\mu+1)(2\mu-1)}{(\mu-1)} \\
\end{array}} \right)
\end{equation}
gives the consistency equation where again the $b$-dependence has cancelled.
Clearly this determinant does not factorize into two distinct closed form
expressions in arbitrary dimensions like the $\lambda_g$ case, except in three
dimensions when one off diagonal element vanishes. Then we would appear to have
$\lambda_{e \, 1}$ $=$ $\eta_1$ $+$ $\chi_{\rho\,1}$ or $\lambda_{e \,1}$ $=$
$0$. For arbitrary dimensions, however, one can obtain expressions for
$\lambda_e$. Although they are the roots of a quadratic equation we do not give
them here for the following reasons.

First, the non-factorization is in fact an indication of a breakdown in the
formalism for {\em this} exponent. To see that it is not a good quantity to
evaluate we repeated its computation in the $O(N)$ Gross Neveu model on its
own and found
\begin{equation}
\lambda^{\mbox{\footnotesize{GN}}} ~=~ \mu -2 - \frac{2(\mu-2)(\mu-1)
\Gamma(2\mu)}{(2\mu-3)\Gamma(2-\mu)\Gamma^2(\mu)\Gamma(\mu+1)N}
\end{equation}
which is clearly singular at $\mu$ $=$ $3/2$. As that model is renormalizable
in the large $N$ expansion in $2$ $\leq$ $d$ $<$ $4$ this exponent is
clearly not a good one to consider and this has been reflected in the
non-factorization of the determinant (5.6) for the amalgam model. In effect,
therefore, what we have learned from this calculation is that for models with
more than one coupling constant the consistency equation will factorize simply
to yield sensible solutions which have a form in keeping with exponents
calculated in arbitrary dimensions by this method in other models. By contrast,
the apparent breakdown of the method in the manner we have discussed here and
represented by non-analytic solutions, is an indication that one is trying
to determine exponents which are inconsistent. Finally, we remark that we have
indeed checked that $\lambda_e$ is $b$-independent for pure QED by deleting
the contributions in (5.6) which come from the $\rho$-terms of (2.1).

\sect{Chern Simons terms.}
Part of our motivation for sudying the theory (2.1) was its potential relation
to models which describe high $T_c$ superconductivity. So far we have
considered the model in fixed arbitrary dimensions and derived various
exponents for all $b$. In this section we consider the three dimensional model
only and extend (2.1) to include a Chern Simons term. We are interested in
determining the effect such a topological term has on the critical exponents
where the coefficient $\theta$ of the Chern Simons term will enter as an
additional parameter. In an abelian gauge theory this coefficient is not
quantized to be an integer as it would be in the non-abelian case, \cite{B}.
Recently, various authors have considered the effect the inclusion of a Chern
Simons term has on the critical exponents of the bosonic $CP(N)$ $\sigma$ model
by conventional renormalization, \cite{13,14}.

Rather than repeat in detail much of the previous discussion, we will
concentrate only those parts of the formalism which are affected by the
inclusion of
\begin{equation}
i \theta \epsilon_{\mu\nu\sigma} A^\mu \partial^\nu A^\sigma
\end{equation}
in (2.1) where $\epsilon_{123}$ $=$ $1$ and $\theta$ is a free parameter. We
recall that we are working in strictly three dimensional euclidean spacetime
which means that the algebra of the $\epsilon$-tensor remains unaltered
compared with what one has to do in explicit perturbative calculations using
dimensional regularization. (We now omit the $(F_{\mu\nu})^2$ term of (2.1). A
transverse contribution for the $A_\mu$ propagator will be generated within the
large $N$ expansion through the relevant graph of fig. 1.) With (6.1) the
asymptotic scaling functions for the gauge field become
\begin{eqnarray}
A_{\mu\nu}(x) & \sim & \frac{C}{(x^2)^\gamma} \left[ \eta_{\mu\nu}
+ \frac{2\gamma(1-b)}{(2-2\gamma+b)} \frac{x_\mu x_\nu}{x^2}
+ i \theta h(\gamma) \frac{\epsilon_{\mu\nu\sigma}
x^\sigma}{(x^2)^{\halfsmall}}
\right. \nonumber \\
\!\!\!&+& \!\!\! \left. C^\prime (x^2)^\lambda \left( \eta_{\mu\nu}
+ \frac{2(\gamma-\lambda) (1-b)}{(2-2\gamma-2\lambda+b)}
\frac{x_\mu x_\nu}{x^2}
+ i \theta h(\gamma-\lambda) \frac{\epsilon_{\mu\nu\sigma}x^\sigma}
{(x^2)^\halfsmall} \right) \right] \nonumber \\
\nonumber \\
A^{-1}_{\mu\nu}(x) & \sim & \frac{m(\gamma,b)}{(1+\theta^2)C
(x^2)^{2\mu-\gamma}} \left[ \eta_{\mu\nu} + \frac{(3-\gamma)}{(\gamma-2)}
\frac{x_\mu x_\nu}{x^2} - i \theta l(\gamma) \frac{\epsilon_{\mu\nu\sigma}
x^\sigma}{(x^2)^\halfsmall} \right. \nonumber \\
\!\! &-& \!\! \left. n(\gamma,b)C^\prime (x^2)^\lambda \! \left( \eta_{\mu\nu}
+ \frac{(3-\gamma-\lambda)}{(\gamma+\lambda-2)} \frac{x_\mu x_\nu}{x^2}
- i \theta l(\gamma+\lambda) \frac{\epsilon_{\mu\nu\sigma}x^\sigma}
{(x^2)^{\halfsmall}} \right) \right] \nonumber \\
\end{eqnarray}
where
\begin{equation}
h(\gamma) ~=~ \frac{2(3-2\gamma)a(1-\gamma)}{(\gamma-1)^2a(\threehalves
-\gamma)} ~~,~~
l(\gamma) ~=~ \frac{(2\gamma-3)a(\gamma-2)}{2(\gamma-2)^2
a(\gamma-\threehalves)} \nonumber
\end{equation}
for an arbitrary covariant gauge. The effect of the presence of (6.1) is to
introduce a factor $(1+\theta^2)^{-1}$ into the $2$-point function. Essentially
this arises from the square of the $\epsilon_{\mu\nu\sigma}k^\sigma$ term in
the inversion on the transverse subspace of momentum space in (2.5). We note
that with the leading order form of (6.2) we can correctly reproduce the
results of \cite{13,14}.

Substituting the leading order terms of (6.2) into the skeleton Dyson
equations of fig. 1 we find
\begin{equation}
\eta_1 ~=~ \frac{2[6b-3+\theta^2]}{3\pi^2(1+\theta^2)}
\end{equation}
from which we determine
\begin{equation}
\chi_{\rho \, 1} ~=~ - \, \frac{2(2b+3-\theta^2)}{(1+\theta^2)\pi^2}
\end{equation}
and $\chi_{A\,1}$ $=$ $- \, \eta_1$ from a vertex renormalization, \cite{15}.
It is worth noting that in the combination $\eta_1$ $+$ $\chi_{\rho \, 1}$ the
gauge parameter cancels to leave a $b$-independent expression for the anomalous
dimension of the $\rho$ field which, here, is a monotonic function of
$\theta^2$, ie $\beta$ $=$ $1$ $-$ $8(\theta^2-3)/ [3\pi^2(1+\theta^2)N]$. To
compute the corrections to $\lambda$ for non-zero $\theta$ we first of all need
to recalculate the $2$-loop graphs of fig. 2 with an internal gauge field as
its propagator has a non-zero $\theta$-dependence. It turns out that the
contribution which is linear in $\theta$ in (6.1) gives zero on integrating the
non-trivial terms which remain after one computes the trace over the
$\gamma$-matrices explicitly using
\begin{equation}
\gamma^\mu \gamma^\nu ~=~ \eta^{\mu\nu} ~+~ i \epsilon^{\mu\nu\sigma}
\gamma_\sigma
\end{equation}
This is consistent with the observation that the exponents must be an even
function of $\theta$. Consequently, the skeleton Dyson equations lead to the
following determinant which contains (5.3)
\begin{equation}
0 ~=~ \det \left(
{\begin{array}{ccc}
1 & \frac{1}{\pi^2} & \frac{4b}{(1+\theta^2)(1+b)\pi^2} \\
1 & - \frac{\lambda_{g \, 1}}{2} + \frac{(\theta^2-15)}{3\pi^2(1+\theta^2)}
& \frac{4}{(1+\theta^2)\pi^2} \\
1 & \frac{1}{\pi^2} & - \, \frac{1}{3(1+b)} \left( \mbox{\small${3\lambda_{g \,
1}}$} - \frac{8b}{(1+\theta^2)\pi^2} \right) \\
\end{array}}
\right)
\end{equation}
which again factorizes into two $b$ independent solutions ie
\begin{equation}
\lambda_{g \, 1} ~=~ 0 ~~~,~~~
\lambda_{g \, 1} ~=~ - \, \frac{4(\theta^2+9)}{3\pi^2(1+\theta^2)}
\end{equation}

We make several remarks on these three dimensional exponents, (6.7). First,
the $\theta$-dependence is as expected non-trivial and they contain the
$\theta$ $=$ $0$ results. Second, when $\theta$ $\rightarrow$ $\infty$ we
find
\begin{equation}
\eta ~ \rightarrow ~ \frac{2}{3\pi^2N} ~~,~~
\chi_\rho ~ \rightarrow ~ \frac{2}{\pi^2N} ~~,~~
\lambda_g ~ \rightarrow ~ \frac{1}{2} - \frac{4}{3\pi^2N}
\end{equation}
However, if we recall that the same critical exponents in the $O(N)$ Gross
Neveu model are at leading order, \cite{4}
\begin{equation}
\eta ~=~ \frac{8}{3\pi^2N} ~~,~~ \chi~=~ \frac{8}{\pi^2N} ~~,~~
\lambda ~=~ \frac{1}{2} - \frac{16}{3\pi^2N}
\end{equation}
we observe that in this $\theta$ $\rightarrow$ $\infty$ limit the critical
exponents for the $O(2N)$ Gross Neveu model are obtained which is a non-trivial
check on our calculation. In this particular limit the gauge field decouples
from the lagrangian leaving only the $4$-fermi interaction. The factor of $4$
discrepancy between the two sets (6.8) and (6.9) is accounted for by the fact
that first we use the convention $\mbox{tr}1$ $=$ $4$ and complex fermions here
as opposed to $\mbox{tr}1$ $=$ $2$ and Majorana fermions in \cite{4}.

\sect{Discussion.}
We conclude with various observations. First, we have now completed a very
comprehensive analysis of the model (2.1) which includes the leading order work
of \cite{15} and next to leading order results of \cite{19} as well as the new
exponents calculated here. Crucially we have included a Chern Simons term in
the formalism of \cite{1,2} which has allowed us to deduce the non-trivial
$\theta$-dependence of gauge independent exponents when topological terms are
included. Whilst much of the discussion was concentrated on $\lambda$ type
exponents, it should not be overlooked that the anomalous dimensions of the
$\rho$ and $A_\mu$ fields are both $b$-independent and therefore physically
relevant. Experimentally the results will be important in allowing one to
discover whether such topological structures are realised in nature or not.
Further, as we now have expressions where $\theta$ appears explicitly in the
exponents it will be possible to tune $\theta$ to particular values in order
that they agree numerically with the same exponents in other models. In other
words they will be in the same universality class as (2.1) for this value of
$\theta$ and therefore this will provide a way of discovering other models
which will be relevant for understanding superconductivity. Indeed the
non-renormalization of $\theta$ to all orders in perturbation theory which has
been proved recently in \cite{23} using algebraic methods is important in this
argument. In other words, $\theta$ remains as a pure parameter quantum
mechanically. Finally, as one of our aims was also to gain a deeper
understanding of how the formalism of \cite{1,2} relates to models with more
than one coupling constant, we record that we have been able to demonstrate
explicitly where one has to be careful in naively applying the method. Further,
we dealt at length with the detailed evaluation of the two loop Feynman
integrals in section 4. The provision of closed expressions for various graphs
is a first step in calculating higher order corrections to $\lambda_{g \, 1}$
which is important for extending the range of $N$ for which reliable estimates
for the exponents can be obtained.
\appendix
\sect{Various recursion relations.}
We summarize several general recursion relations, in the notation of
section 4, which were required to evaluate the two loop Feynman graphs. They
were constructed using an integration rule similar to fig. 3 but where the
sum of the exponents is $2\mu$ $+$ $1$. Thus,
\begin{eqnarray}
\la \alpha_1, \alpha_2, \alpha_3, \alpha_4, \alpha_5 \ra &=&
\frac{(2\mu-t_1)(t_1-\mu)}{(\alpha_2-1)(\alpha_3-1)} [0,-1,-1,0,1]_+
\nonumber \\
&+& \frac{(\alpha_2+\alpha_3-\mu-1)}{(\alpha_2-1)} [0,-1,0,-1,+1]_+
\nonumber \\
&+& \frac{(\alpha_2+\alpha_3-\mu-1)}{(\alpha_3-1)} [-1,0,-1,0,+1]_+ \\
\la \alpha_1, \alpha_2, \alpha_3, \alpha_4, \alpha_5 \ra &=&
\frac{\alpha_2\alpha_5}{(\alpha_1-1)(\mu-\alpha_1)} [-1,+1,-1,0,+1] \nonumber
\\
&+& \frac{\alpha_5(\mu-\alpha_2-\alpha_5-1)}{(\alpha_1-1)(\mu-\alpha_1)}
[-1,0,0,0,+1] \nonumber \\
&+& \frac{\alpha_2(\mu-\alpha_2-\alpha_5-1)}{(\alpha_1-1)(\mu-\alpha_1)}
[-1,+1,0,0,0] \\
\la \alpha_1, \alpha_2, \alpha_3, \alpha_4, \alpha_5 \ra &=&
\frac{(d-2\mu-1)(3\mu-d)}{(\alpha_2-1)(\alpha_3-1)} [0,-1,-1,0,1]_+ \nonumber
\\
&+& \frac{(\alpha_2+\alpha_3-\mu-1)}{(\alpha_2-1)} [0,-1,0,0,1] \nonumber \\
&+& \frac{(\alpha_2+\alpha_3-\mu-1)}{(\alpha_3-1)} [0,0,-1,0,1]
\end{eqnarray}
where we have defined $[0,0,1,-1,0]$ to mean, for example, $\la \alpha_1,
\alpha_2, \alpha_3+1, \alpha_4-1, \alpha_5 \ra$. The appearance of a $\pm$ sign
subscript on this bracket indicates that there is an overall factor of
$(x^2)^{\mp 1}$ multiplying the two loop graph. Also, we have introduced the
following linear combinations of $\alpha_i$
\begin{eqnarray}
t_1 ~=~ \alpha_1 + \alpha_4 + \alpha_5 &,& t_2 ~=~ \alpha_2 + \alpha_3
+ \alpha_5 \nonumber \\
s_1 ~=~ \alpha_1 + \alpha_2 + \alpha_5 &,& s_2 ~=~ \alpha_3 + \alpha_4
+ \alpha_5 \nonumber \\
\! d ~=~ \sum_{i=1}^{5} \alpha_i ~~~~~~~~~~~\!\! &&
\end{eqnarray}
\sect{Various finite integrals.}
In this appendix, we record the various finite integrals which we needed to
calculate the graphs of fig. 2 when broken up into their consistuent bosonic
pieces. They were derived using recursion relations similar to those of
appendix A and the methods of \cite{21},
\begin{eqnarray}
&&\la \mu-1, \mu-1, \mu-1, \mu-1, 4-\mu \ra ~=~ \frac{\pi^{2\mu}(\mu-1)^2}
{(\mu-2)^2\Gamma^2(\mu)} \\
&&\la \mu-2, \mu-1, \mu-1, \mu-1, 4-\mu \ra \nonumber \\
&& ~~=~ - \, \frac{a(4-\mu)}{a^3(1)} \left[ \frac{\pi^{2\mu}\Gamma(\mu-1)
\Gamma(2-\mu)}{\Gamma(2\mu-3)} + (\mu-2)^2 ChT(1,1) \right] \\
&&\la \mu-1, \mu-1, \mu-1, \mu-1, 5-\mu \ra ~=~ \frac{\pi^{2\mu} (\mu-1)^3}
{2(\mu-3)(\mu-2)^2 \Gamma^2(\mu)} \\
&& \la \mu-2, \mu-1, \mu-1, \mu-1, 5-\mu \ra ~=~ \frac{\pi^{2\mu} (\mu-1)^2}
{2(\mu-2)(\mu-3)\Gamma^2(\mu)} \\
&&\la \mu-3, \mu-1, \mu-1, \mu-1, 5-\mu \ra \nonumber \\
&& ~~=~ \frac{a(5-\mu)}{a(1)a^2(2)} \left[ (\mu-3)^2 ChT(1,1)
+ \frac{\pi^{2\mu}(\mu^2-7\mu+11)a(1)}{(\mu-2)^3a(3-\mu)}
\right] \\
&&\la \mu-2, \mu-1, \mu-1, \mu-2, 5-\mu \ra \nonumber \\
&& ~~=~ \frac{a(5-\mu)(\mu-2)}{a^3(2)} \left[ ChT(1,1) - \frac{\pi^{2\mu}
(\mu-3)a(1)}{(\mu-2)^3a(3-\mu)} \right] \\
&& \la \mu-2, \mu-1, \mu-2, \mu-1, 5-\mu \ra \nonumber \\
&& ~~=~ \frac{a(5-\mu)}{a^3(2)} \left[ ChT(1,1)
+ \frac{\pi^{2\mu}(2\mu^2 - 12\mu + 17)a(2)}{(\mu-2)^2a(3-\mu)} \right] \\
&& \la \mu-1, \mu-1, \mu-1, \mu-1, 6-\mu \ra \nonumber \\
&& ~~~~~~~~~~~~~~~~~~~=~~ \frac{\pi^{2\mu} (\mu-1)^2 (5\mu^3 - 27 \mu^2
+ 40 \mu -24)}{18 (\mu-2)^2(\mu-3)(\mu-4)\Gamma^2(\mu)} \\
&& \la \mu-2, \mu-1, \mu-1, \mu-1, 6-\mu \ra \nonumber \\
&& ~~~~~~~~~~~~~~~~~~~=~~ \frac{\pi^{2\mu} (\mu-1)^2 (3\mu^2 - 14 \mu + 12)}
{12(\mu-2)(\mu-3)^2(\mu-4) \Gamma^2(\mu)} \\
&& \la \mu-2, \mu-1, \mu-2, \mu-1, 6-\mu \ra ~=~ \! \frac{\pi^{2\mu}(\mu-4)
(\mu-1)^2}{4(\mu-2)(\mu-3)^2 \Gamma^2(\mu)} ~~~~ \\
&& \la \mu-2, \mu-1, \mu-1, \mu-2, 6-\mu \ra ~=~ \! \frac{\pi^{2\mu}(\mu-1)^2}
{6(\mu-3)(\mu-4) \Gamma^2(\mu)} \\
&& \la \mu-3, \mu-1, \mu-1, \mu-1, 6-\mu \ra ~=~ \! \frac{\pi^{2\mu} (\mu-1)^2}
{3(\mu-2)(\mu-4) \Gamma^2(\mu)} ~~~~
\end{eqnarray}
The integral $ChT(1,1)$ appears in several of the basic integrals above. It is
a special case of the integral $ChT(\alpha,\beta)$, in the notation of
\cite{2},
which is defined as $\la \alpha ,\mu-1,\mu-1,\beta,\mu-1 \ra$, for arbitrary
$\alpha$ and $\beta$. It has been evaluated exactly in, for example, \cite{2}.
For the particular case we have $ChT(1,1)$ $=$ $3\pi^{2\mu} a(1)a(2\mu-1)
[\psi^\prime(\mu-1) - \psi^\prime(1)]$, where $\psi(\mu)$ is the logarithmic
derivative of the $\Gamma$-function.

Next, we give the explicit expressions for each finite linear combination of
divergent graphs defined in sect. 4, for arbitrary $\xi$.
\begin{eqnarray}
B_\xi &=& \frac{(4\alpha+\xi-2\mu-3)(3\mu-4\alpha-\xi+2)}{(\alpha-1)^4}
\left[ \frac{(\mu-1)\xi(\xi-\mu+1)}{(2\alpha+\xi-\mu-1)} \right. \nonumber \\
&& +~ \left. 2(\mu-\alpha)^2 \frac{}{} \right] \la \alpha-1, \alpha-1,
\alpha-1, \alpha-1, \xi+1 \ra \nonumber \\
&+& \frac{2(\mu-\alpha)}{(\alpha-1)^3} \left[ 2(\mu-\alpha)(2\alpha-\mu
-1) - \frac{(\mu-1)\xi(\xi-\mu+1)}{(2\alpha+\xi-\mu-1)} \right] \nonumber \\
&&\times ~\la \alpha-1, \alpha-1, \alpha, \alpha-1, \xi+1 \ra
\end{eqnarray}
\begin{eqnarray}
C_\xi &=& \frac{(4\alpha+\xi-2\mu-4)(3\mu-4\alpha-\xi+3)}{(\alpha-1)^2}
\la \alpha-1, \alpha-1, \alpha-1, \alpha-1, \xi \ra \nonumber \\
&-& \frac{\xi(\mu-\xi-1)}{(\alpha-1)^2} \la \alpha-1, \alpha-1, \alpha-1,
\alpha-1, \xi+1 \ra
\end{eqnarray}
\begin{eqnarray}
D_\xi &=& - \, \frac{2(\mu-\alpha)(\alpha+\xi-2)}{(\alpha-1)(\alpha-2)}
\la \alpha-1, \alpha-1, \alpha, \alpha-1, \xi \ra \nonumber \\
&+& \frac{\xi}{(\alpha-1)} \left[ 2 + \frac{(2\alpha+\xi-\mu-3)
(\mu-\alpha-\xi)}{(\alpha-1)(\alpha-2)} \right] \nonumber \\
&&\times~ \la \alpha-1, \alpha-1, \alpha-1, \alpha-1, \xi+1 \ra \nonumber \\
&+& \frac{2(\mu-\alpha)(2\mu-2\alpha-\xi+2)(2\alpha+\xi-\mu-3)}
{(\alpha-1)(\alpha-2)(2\alpha+\xi-\mu-2)(2\mu-2\alpha-\xi+1)} \nonumber \\
&&\times~ \la \alpha-1, \alpha-1, \alpha, \alpha-1, \xi-1 \ra \nonumber \\
&+& f(\alpha, \xi) \, \la \alpha-1, \alpha-1, \alpha-1, \alpha-1,
\xi \ra
\end{eqnarray}
\begin{eqnarray}
E_\xi &=& \frac{\xi(\mu-\xi-1)}{(\alpha-1)^2} \la \alpha-1, \alpha-1,
\alpha-1, \alpha-1, \xi+1 \ra \nonumber \\
&-& \frac{2(\mu-\alpha)}{(\alpha-1)} \la \alpha-1, \alpha-1, \alpha,
\alpha-1, \xi \ra
\end{eqnarray}
where we have set
\begin{eqnarray}
f(\alpha, \xi) \! &=& \! \! \frac{(3\mu-4\alpha-\xi+3)(\mu-\alpha-\xi+1)
(2\mu-2\alpha-\xi+2)}{(\alpha-1)^2(\alpha-2)(2\mu-2\alpha-\xi+1)
(2\alpha+\xi-\mu-2)} \nonumber \\
&\times& \! \! [ \frac{}{} (\alpha-1)(2\mu-2\alpha-\xi+1)
(4\alpha+2\xi-3\mu-4)(2\alpha+\xi-4) ~~~~~~ \nonumber \\
&+& (\mu-\alpha-\xi+1)(2\mu-2\alpha-\xi+2) \nonumber \\
&&\!\!\!\!\!\! \times \,\, [(\mu-\xi)(2\alpha+\xi-\mu-3)
- (4\alpha+2\xi-3\mu-4)(\mu-1)]]
\end{eqnarray}

\newpage
\noindent
{\Large {\bf Figure Captions.}}
\begin{description}
\item[Fig. 1.] Leading order dressed skeleton Dyson equations.
\item[Fig. 2.] Additional graphs contributing to consistency equations.
\item[Fig. 3.] Uniqueness rule for a bosonic vertex.
\item[Fig. 4.] General two loop self energy.
\item[Fig. 5.] General integration by parts rule for a gauge vertex.
\end{description}
\end{document}